\documentclass[aps,prl,a4paper,superscriptaddress,amsmath,amssymb,twocolumn]{revtex4} 
\usepackage{graphicx,epsfig}
\usepackage[usenames]{color}

\allowdisplaybreaks
\begin{document}

\newcommand{\Avg}[1]{\langle \,#1\,\rangle_G}

\newcommand{\E}{\mathcal{E}}
\newcommand{\Lag}{\mathcal{L}}
\newcommand{\M}{\mathcal{M}}
\newcommand{\N}{\mathcal{N}}
\newcommand{\U}{\mathcal{U}}
\newcommand{\R}{\mathcal{R}}
\newcommand{\F}{\mathcal{F}}
\newcommand{\V}{\mathcal{V}}
\newcommand{\C}{\mathcal{C}}
\newcommand{\I}{\mathcal{I}}
\newcommand{\s}{\sigma}
\newcommand{\up}{\uparrow}
\newcommand{\dw}{\downarrow}
\newcommand{\h}{\hat{\mathcal{H}}}
\newcommand{\Hn}{\mathcal{H}}
\newcommand{\himp}{\hat{h}}
\newcommand{\g}{\mathcal{G}^{-1}_0}
\newcommand{\D}{\mathcal{D}}
\newcommand{\A}{\mathcal{A}}
\newcommand{\projs}{\hat{\mathcal{S}}_d}
\newcommand{\proj}{\hat{\mathcal{P}}_d}
\newcommand{\K}{\textbf{k}}
\newcommand{\Q}{\textbf{q}}
\newcommand{\hzero}{\hat{T}}
\newcommand{\io}{i\omega_n}
\newcommand{\eps}{\varepsilon}
\newcommand{\+}{\dag}
\newcommand{\su}{\uparrow}
\newcommand{\giu}{\downarrow}
\newcommand{\0}[1]{\textbf{#1}}
\newcommand{\ca}{c^{\phantom{\dagger}}}
\newcommand{\cc}{c^\dagger}
\newcommand{\aaa}{a^{\phantom{\dagger}}}
\newcommand{\aac}{a^\dagger}
\newcommand{\bba}{b^{\phantom{\dagger}}}
\newcommand{\bbc}{b^\dagger}
\newcommand{\da}{d^{\phantom{\dagger}}}
\newcommand{\dc}{d^\dagger}
\newcommand{\fa}{f^{\phantom{\dagger}}}
\newcommand{\fc}{f^\dagger}
\newcommand{\ha}{h^{\phantom{\dagger}}}
\newcommand{\hc}{h^\dagger}
\newcommand{\be}{\begin{equation}}
\newcommand{\ee}{\end{equation}}
\newcommand{\bea}{\begin{eqnarray}}
\newcommand{\eea}{\end{eqnarray}}
\newcommand{\ba}{\begin{eqnarray*}}
\newcommand{\ea}{\end{eqnarray*}}
\newcommand{\dagga}{{\phantom{\dagger}}}
\newcommand{\bR}{\mathbf{R}}
\newcommand{\bQ}{\mathbf{Q}}
\newcommand{\bq}{\mathbf{q}}
\newcommand{\bqp}{\mathbf{q'}}
\newcommand{\bk}{\mathbf{k}}
\newcommand{\bh}{\mathbf{h}}
\newcommand{\bkp}{\mathbf{k'}}
\newcommand{\bp}{\mathbf{p}}
\newcommand{\bL}{\mathbf{L}}
\newcommand{\bRp}{\mathbf{R'}}
\newcommand{\bx}{\mathbf{x}}
\newcommand{\by}{\mathbf{y}}
\newcommand{\bz}{\mathbf{z}}
\newcommand{\br}{\mathbf{r}}
\newcommand{\Ima}{{\Im m}}
\newcommand{\Rea}{{\Re e}}
\newcommand{\Pj}[2]{|#1\rangle\langle #2|}
\newcommand{\ket}[1]{\vert#1\rangle}
\newcommand{\bra}[1]{\langle#1\vert}
\newcommand{\setof}[1]{\left\{#1\right\}}
\newcommand{\fract}[2]{\frac{\displaystyle #1}{\displaystyle #2}}
\newcommand{\Av}[2]{\langle #1|\,#2\,|#1\rangle}
\newcommand{\av}[1]{\langle #1 \rangle}
\newcommand{\Mel}[3]{\langle #1|#2\,|#3\rangle}
\newcommand{\Avs}[1]{\langle \,#1\,\rangle_0}
\newcommand{\eqn}[1]{(\ref{#1})}
\newcommand{\Tr}{\mathrm{Tr}}

\newcommand{\Pg}{\mathcal{P}_G}

\newcommand{\Vb}{\bar{\mathcal{V}}}
\newcommand{\Vd}{\Delta\mathcal{V}}
\newcommand{\Pb}{\bar{P}_{02}}
\newcommand{\Pd}{\Delta P_{02}}
\newcommand{\tb}{\bar{\theta}_{02}}
\newcommand{\td}{\Delta \theta_{02}}
\newcommand{\Rb}{\bar{R}}
\newcommand{\Rd}{\Delta R}

\title{The $\gamma$-$\alpha$ iso-structural Transition in Cerium, a Critical Element}

\author{Nicola Lanat\`a}
\altaffiliation{Equally contributed to this work}
\affiliation{Department of Physics and Astronomy, Rutgers University, Piscataway, New Jersey 08856-8019, USA}
\author{Yong-Xin Yao}
\altaffiliation{Equally contributed to this work}
\affiliation{Ames Laboratory-U.S. DOE and Department of Physics and Astronomy, Iowa State
University, Ames, Iowa IA 50011, USA}
\author{Cai-Zhuang Wang}
\affiliation{Ames Laboratory-U.S. DOE and Department of Physics and Astronomy, Iowa State
University, Ames, Iowa IA 50011, USA}
\author{Kai-Ming~Ho}
\affiliation{Ames Laboratory-U.S. DOE and Department of Physics and Astronomy, Iowa State
University, Ames, Iowa IA 50011, USA}
\author{J\"org Schmalian}
\affiliation{Karlsruhe Institute of Technology, Institute for Theory of Condensed Matter, D-76131 Karlsruhe, Germany}
\author{Kristjan Haule}
\affiliation{Department of Physics and Astronomy, Rutgers University, Piscataway, New Jersey 08856-8019, USA}
\author{Gabriel Kotliar}
\affiliation{Department of Physics and Astronomy, Rutgers University, Piscataway, New Jersey 08856-8019, USA}

\date{\today} 

\begin{abstract}

{\bf Below the critical temperature $T_c\simeq 600 K$, 
an iso-structural transition, named $\gamma$-$\alpha$ transition,
can be induced in Cerium by applying pressure. 
This transition is first-order, and is accompanied by a sizable
volume collapse. 
A conclusive theoretical explanation of this intriguing
phenomenon has still not been achieved, and the 
physical
pictures proposed so far 
are 
still under debate.
In this work, we illustrate 
zero-temperature first-principle
calculations which clearly demonstrate that
the $\gamma$-$\alpha$ transition is
induced by the  interplay between 
the electron-electron Coulomb interaction
and the spin-orbit coupling. 
We address the still unresolved problem on the existence of 
a second low-$T$ critical point, i.e.,
whether the energetic effects alone are sufficient or not
to induce the $\gamma$-$\alpha$ transition 
at zero temperature.
}

\end{abstract}

\maketitle

The $\gamma$-$\alpha$ iso-structural transition in Cerium~\cite{Gschneidner78}
was discovered in 1949~\cite{Lawson49}.
Since then, a lot of theoretical and experimental work has been
devoted to its understanding.
The great interest in this phenomenon arises from the 
fact that the transition is isostructural,
i.e., the lattice structure of the system is equal in the two phases.
Furthermore, the possibility that 
the underlying mechanism lies in the electronic structure only ---
i.e., without it being necessary to involve other effects ---
makes Cerium a potential theoretical
testing ground for basic concepts of correlated electron systems.
Two main theoretical pictures are still under debate 
to explain the volume collapse: 
the Kondo volume collapse (KVC)~\cite{Allen82,Lavagna82} and the 
orbital-selective Mott transition within the
Hubbard model (HM)~\cite{Johansson74}.
According to the KVC the transition is induced by the rapid 
change of the coherence temperature
across the transition boundaries, which affect dramatically 
the structure of the conduction $spd$ electrons through Kondo effect. 
According to the HM, instead, it is
the hopping between $f$ orbitals that changes drastically
across the transition between the $\alpha$ phase (with delocalized
$f$ electrons) and the $\gamma$ phase (with localized $f$ electrons),
as for the Mott transition in the Hubbard model.

Consistently with both the HM and the KVC pictures,
the $f$-electrons are strongly correlated
both in the $\alpha$ and in the $\gamma$ phase.
This fact is clearly indicated, e.g., by the
photoemission spectra, which is known 
experimentally~\cite{Wieliczka,Weschke,Patthey}
and theoretically~\cite{Keller01,Held01,McMahan}.
Despite this similarity, there is a key difference between these two models: 
while the KVC attributes a very important
role to the interplay between the localized 4$f$ orbitals
and the itinerant $spd$ conduction bands, the 
itinerant electrons are 
``spectators'' in the HM picture.

The development of LDA+DMFT
(Local Density Approximation plus Dynamical Mean Field Theory)~\cite{LDA+U+DMFT}
results~\cite{Gabi05,Gabi06,Moore09} 
has successfully reproduced many aspects of this transition,
and different aspects of these studies can be understood in both physical pictures.
There are  still  fundamental questions which 
have not been answered. 
(1) What is the role of the spin-orbit interaction (SOC) for the volume-collapse?
(2) What is the fate of the pressure-temperature transition-line
at very low temperatures~\cite{deMedici, Georges06, Scheffler12}?
(3) Can a first-principle based theory be  made computationally efficient
so as to access both the $\gamma$ and the 
$\alpha$ Cerium at zero temperature?

Due to the complexity of the problem, it is clear that, in order
to be conclusive, a theoretical
explanation of the $\gamma$-$\alpha$ transition
needs to be supported by first-principles
calculations which, not only are able to take into account both the details
of the band-structure and the strong-correlation effects, but are also
able to 
evaluate precisely the pressure-volume phase diagram.
For this purpose, 
it is crucial that the computation of the total energy is
essentially free of numerical error.
Another key requirement is that the two phases are treated within the 
same theoretical framework.
In this work, we use a combination of Density Functional Theory
and the Gutzwiller Approximation (LDA+GA)~\cite{Gutz65,Zein,Deng_LDA+Gutz,Ho,Gmethod}, 
which satisfies all of these requirements. 
Recently, we have established formally~[L.N. et al.]   
that this method can be viewed as 
an instance of LDA+DMFT, using Slave Bosons (or the Gutzwiller method)
as the DMFT impurity solver~\cite{Gabi06}.
This insight enabled a new efficient charge self-consistent
implementation of the LDA+GA method
on top~[Y.Y.X. et al.]   
of the LAPW DFT code Wien2k~\cite{w2k}, which removes many of the approximations
inherent in previous studies. 
As a benchmark, in the supplementary material we present also
LDA+DMFT calculations for Cerium.
The very good agreement between the two methods
gives us further confirmation that the results presented in
this work are indeed reliable.

We employ the general Slater-Condon parametrization of the
on-site interaction, assuming a Hund's coupling constant $J=0.7\,eV$~\cite{cowan}.
Since the value $U$ of the interaction strength 
is generally difficult to establish accurately
(due to it's strong sensitivity to the screening effect), in this work
we perform calculations scanning different values of $U$.
Our calculations are all performed at zero temperature.

\begin{figure}
\begin{center}
\includegraphics[width=8.4cm]{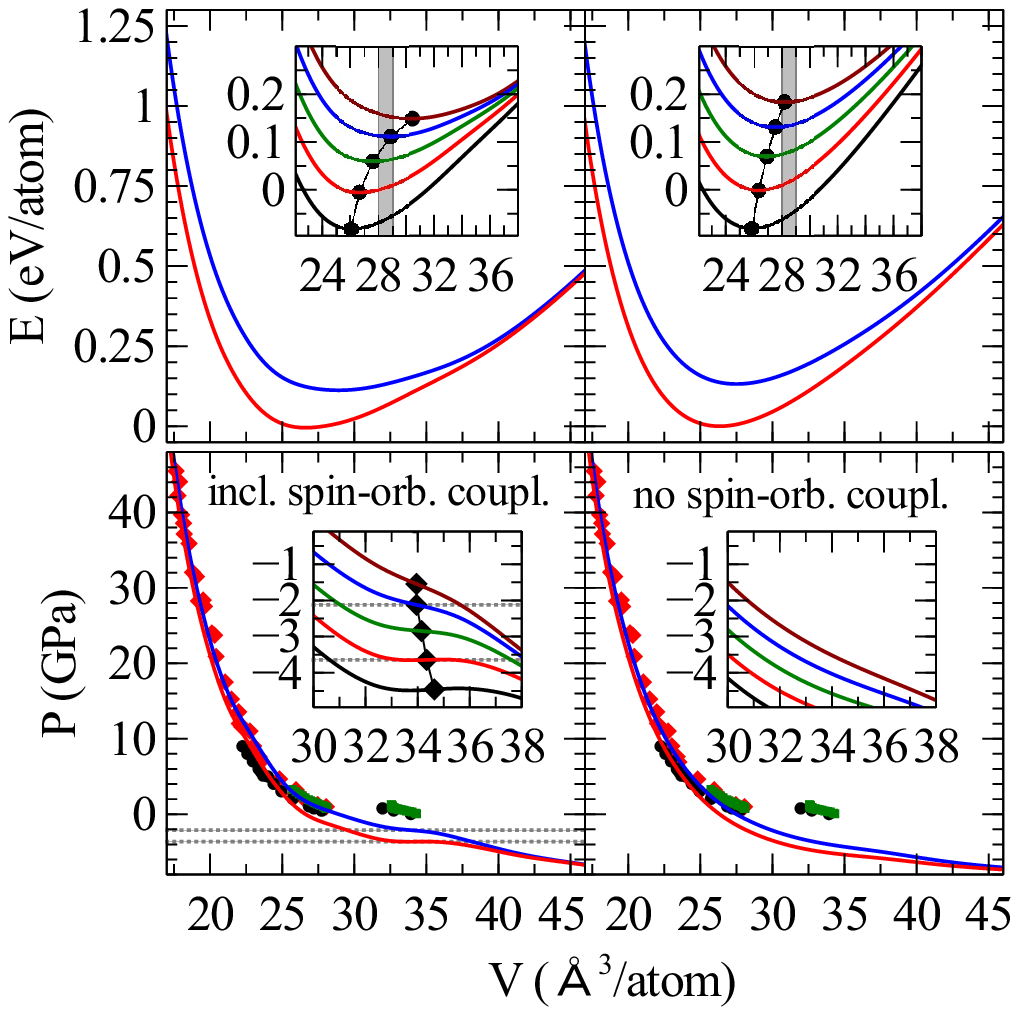}
\caption{Total energy as a function of the volume (upper panels) and
  corresponding theoretical pressure-volume curves for $U=5,6\,eV$, $J=0.7\,eV$,
  in comparison with the experimental data (lower panels).
  The experimental data
  relate to measurements at room temperature 
  (black circles from Ref.~\cite{CePV77},
  red diamonds from Ref.~\cite{CePV85} and
  green squares from Ref.~\cite{Lipp12}), 
  while our theoretical calculations are all obtained at zero temperature.
  The curves in the insets are obtained for all
  $U$'s from $4.5\,eV$ to $6.5\,eV$ (from lower to higher total energies)
  with step of $0.5\,eV$. 
  Our results are shown both with (left panels) 
  and without (right panels) taking into account the spin-orbit coupling.
  The vertical shaded line in the upper insets indicate the experimental
  volume at ambient pressure. 
  The horizontal dotted lines in the lower-left panel
  and the black diamonds in its inset indicate the 
  pressures where the bulk-modulus
  $\mathcal{K}=-V dP/dV$ is minimum.
}

\label{figure1}
\end{center}
\end{figure}

In the upper panels of Fig.~\ref{figure1} we illustrate our theoretical
energy-volume diagrams for different values of $U$. 
The results are shown both 
by taking into account the SOC 
(left panels) and by neglecting it in the calculation (right panels). 
The corresponding pressure-volume curves,
obtained from $P=-dE/dV$,
are shown in the lower panels,
in comparison with the experimental data at room temperature
of Refs.~\cite{CePV77,CePV85}. 
The agreement with the experiment is 
good, especially for $U=6\,eV$,
which is the value that reproduces the 
experimental equilibrium-volume
$V_{\text{eq}}\simeq 28.5\,\AA^3/\text{atom}$~\cite{CePV74,CePV77,CePV85}
(see the inset of the higher-left panel).
The small discrepancies
at larger $V$ are likely, at least in part, due to the entropy, as our calculations
were performed at zero temperature.
Note that $U=6\,eV$ was also previously computed within
the constrained LDA method~\cite{McMahan98,Keller01}, which gives us further 
confidence that this is
the optimal value of the correlation strength for Cerium.

Remarkably, we observe a change of sign in the bulk-modulus 
$\mathcal{K}=-V dP/dV<0$ ---  which is the 
signal of a first-order iso-structural transition --- 
for any $U\leq U_c\simeq 5.5\,eV$
(see the pressure-volume curves in the lower-panels insets of Fig.~\ref{figure1}); 
while at $U=U_c$ the transition becomes
second-order, with $\mathcal{K}=0$ minimum bulk-modulus. 
A signature of the transition, i.e., 
a local minimum of the bulk-modulus, 
is still present for any interaction strength, see the black diamonds
in the inset of the lower-left panel of Fig.~\ref{figure1}.
We point out that this feature of the pressure-volume curve is observed \emph{only}
if the SOC 
is taken into account --- which is a clear
indication of its key role in the physics underlying the $\gamma$-$\alpha$
transition.
Note also that, for $U=6\,eV$, the crossover point occurs at $P\simeq -2\,\text{Gp}$, which is 
close to the expected zero-temperature value $P_{\text{exp}}\simeq -1\,\text{Gp}$
extrapolated from the experimental data 
of Ref.~\cite{Gschneidner78}.

\begin{figure}
\begin{center}
\includegraphics[width=8.4cm]{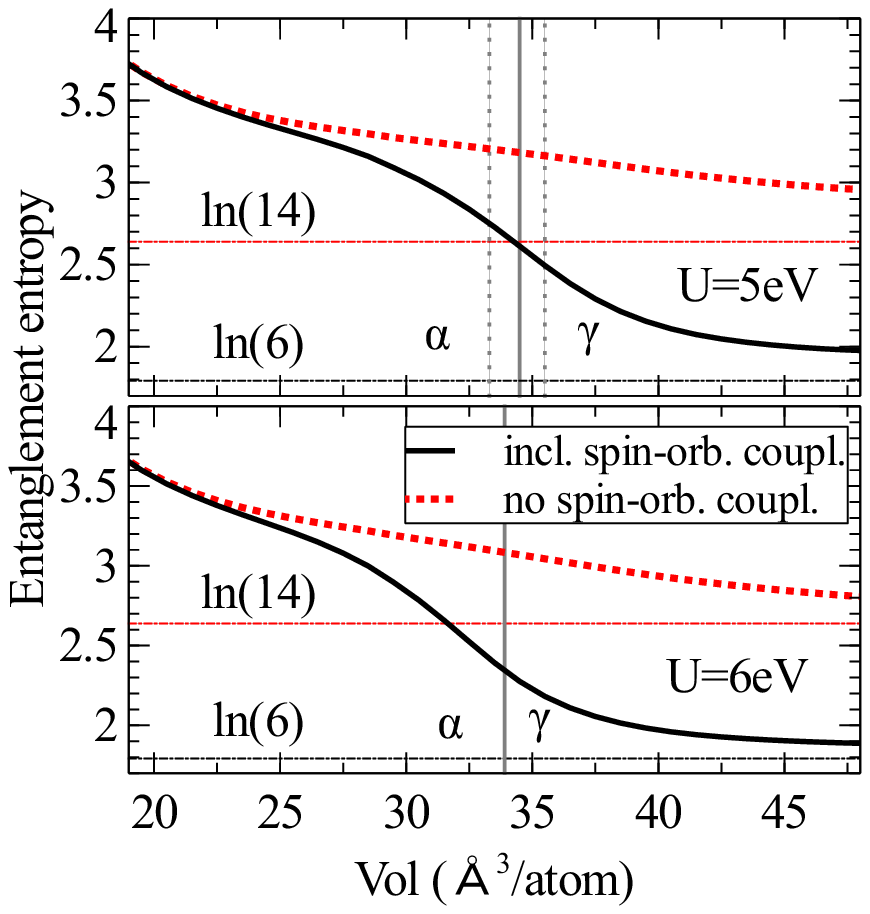}
\caption{Local entanglement entropy of the $f$-electrons
as a function of the volume per atom of the system for $U=5\,eV$ (upper panel)
and $U=6\,eV$ (lower panel) at fixed $J=0.7\,eV$.
The entanglement entropy is reported both for the case with (lines) and without (dots)
the SOC. 
The horizontal lines correspond to
$14$, which is the dimension of the single-particle local space
of the $f$-electrons,
and to $6=2\times 5/2+1$, which is the degeneracy of the $5/2$ $f$-electrons
within the single-particle local space.
The vertical continue lines indicate the signature of the transition
in the pressure-volume diagram, and the dotted vertical lines 
indicate the boundary of the $\gamma$-$\alpha$ transition
(which occurs only for $U\leq U_c\simeq 5.5\,eV$)
according to the equal-area construction~\cite{Landau}.
}
\label{figure2}
\end{center}
\end{figure}

In order to better understand the role of the SOC, 
we consider the local $f$ entanglement entropy,
\be
S_f[\rho_f]=-\Tr\!\left[\rho_f\ln\rho_f\right]\,,
\ee
where $\rho_f$ is the reduced density matrix of the system in the 
$f$ local subspace.
The value of $S_f$ is a measure of how much the $f$ electrons
are entangled with the rest of the environment.
In Fig.~\ref{figure2} the behaviour of $S_f$ is shown as a function
of the volume for two values of $U$.
Remarkably, if (and only if) the SOC 
is taken into account, a clear crossover is visible 
in correspondence of the signature of the volume collapse, 
i.e., in correspondence of the 
minimum of the bulk-modulus 
$\mathcal{K}$, which is indicated by black diamonds
in the inset of the lower-left panel of Fig.~\ref{figure1}.

In the $\alpha$ phase, as expected, 
$S_f$ is not sensitive to the spin-orbit splitting, indicating that the 
local fluctuations
induced in the $f$ local space by
the coupling with its environment are very large.
By increasing the volume, the fluctuations between the $J=5/2$ $f^1$ subspace 
and the other local configurations are increasingly suppressed.
The crossover point 
identifies the situation in which the above-mentioned fluctuations 
are sufficiently small
to be hampered by the spin-orbit splitting.
This is clearly demonstrated by the fact that in the $\gamma$ phase, 
when the SOC is taken into account, $S_f\gtrsim\ln 6$
--- 
where $6=2\times 5/2+1$ is the degeneracy of the $5/2$ eigenspace at $n_f=1$.
We point out that the above-mentioned
local fluctuations are generated
only by the entanglement, and so are present even if the actual temperature
of the system is zero --- as  
in our calculations.
As we are going to show, 
the main source of entanglement is the hybridization between the
$f$ and the $spd$ electrons.


\begin{figure}
\begin{center}
\includegraphics[width=8.4cm]{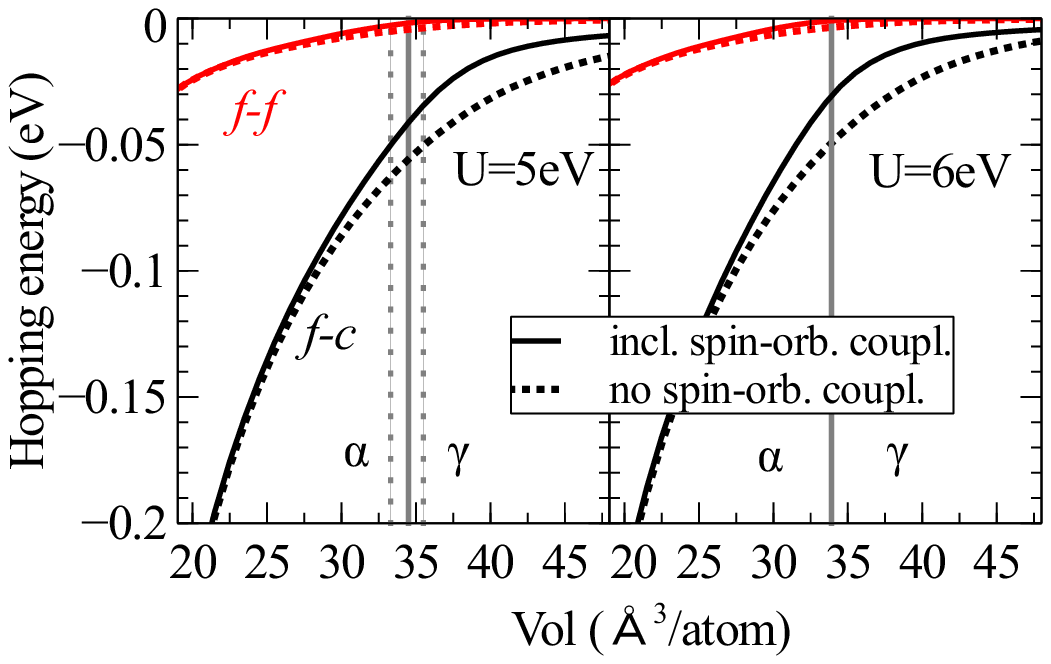}
\caption{Ground-state expectation values of the $f$-$c$ and the $f$-$f$
effective hopping energies
as a function of the volume per atom of the system.
The energies are reported for $U=5\,eV$ (left panel)
and $U=6\,eV$ (right panel) at fixed $J=0.7\,eV$, both for the case with (lines) 
and without (dots) the spin-orbit coupling.
The vertical continue lines indicate the signature of the transition
in the pressure-volume diagram, and the dotted vertical lines 
indicate the boundary of the $\gamma$-$\alpha$ transition
(which occurs only for $U\leq U_c\simeq 5.5\,eV$)
according to the equal-area construction~\cite{Landau}.
}
\label{figure3}
\end{center}
\end{figure}

A further insight of the problem can be achieved by inspecting
the ground-state expectation values of the 
non-local energy components of the effective 
Hamiltonian $\h$
whose ground-state provides our theoretical solution~\cite{Deng_LDA+Gutz,LDA+U+DMFT}.
The non-local part $\hat{T}$ of $\h$ 
can be concisely represented as
\be
\hat{T}=\hzero_{ff}+\hzero_{fc}+\hzero_{cc}\,, 
\ee
where the symbol $c$ represents all of the $spd$ conduction electrons, 
and
$\hzero_{ff}$, $\hzero_{fc}$ and $\hzero_{cc}$ represent
the non-local ``hopping'' terms between $f$-$f$, $f$-$c$ and $c$-$c$
electrons, respectively.
In Fig.~\ref{figure3} the ground-state expectation values of the
$f$-$c$ and the $f$-$f$  
components of $\hat{T}$  
are shown for two values of $U$. 
These energies represent the Kondo and the Hubbard energy scales
of the problem, respectively.
In agreement with the KVC model of the transition, we observe that
the Kondo energy scale is about one order of magnitude bigger
than the Hubbard energy scale, which is already very small before
the crossover point, as expected~\cite{Taylor05}.
This confirms that the main source of entanglement between the $f$ local
space and its environment is the 
hybridization between the $f$ and the $spd$ electrons.
Note that, when the SOC is taken into account, 
a more rapid suppression of the Kondo energy scale is observed concomitantly with
the crossover region.

We have already observed that,
even though the behaviour of the entanglement entropy 
(and of the Kondo energy scale) are qualitatively the same for all $U$'s, 
no transition can be found for $U\geq U_c\simeq 5.5\,eV$ at zero temperature,
see Figs.~\ref{figure1} and \ref{figure2}. 
The reason is the following. The local crossover, 
which induces a reduction of the bulk-modulus 
of the system,
occurs at lower volumes for larger $U$, see the black diamonds
in the inset of the lower-left panel of Fig.~\ref{figure1}.
On the other hand, the bulk-modulus  
becomes larger at smaller volumes
(even when the SOC is not taken into account),
see the pressure-volume curves in the lower panels of Fig.~\ref{figure1}.
For this reason, if $U$ is large enough, it becomes impossible for the
SOC to make the bulk-modulus 
negative,
i.e., to induce the volume collapse.
It follows that, in principle, there are two possible scenarios:
(i) the $\gamma$-$\alpha$ transition exists also at zero temperature, 
or (ii) the transition line ends at a certain finite
critical temperature (at negative pressures).
As we mentioned before,
$U\simeq 6\,eV$ --- which is indeed very close to $U_c$ ---
is a physically reasonable value for Cerium. 
This suggests that Cerium is placed essentially in the middle between
the two above-mentioned scenarios, i.e., that the $\gamma$-$\alpha$ 
transition line ends very close to zero temperature.
Note that this finding is in qualitative agreement with the experimental results
of Ref.~\cite{Lashley}.


\begin{figure}
\begin{center}
\includegraphics[width=8.4cm]{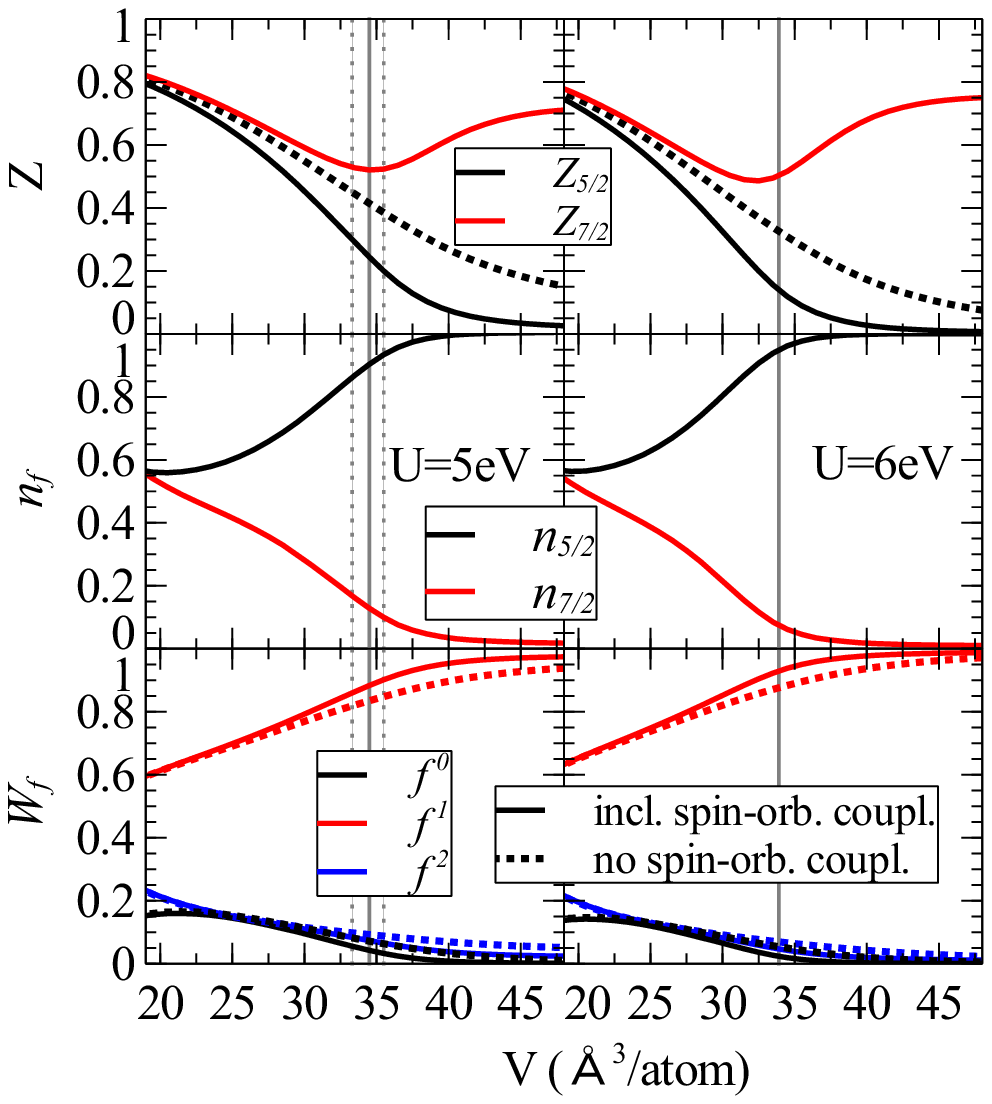}
\caption{Quasi-particle renormalization weights
  of the $7/2$ and $5/2$ $f$-electrons  (upper panels),
  $7/2$ and $5/2$ $f$ orbital populations (central panels),
  and $f$ configuration probabilities (lower panels),
  as a function of the volume of the system.
  The renormalization weights and the configuration probabilities
  are reported both for the case with 
  and without the spin-orbit coupling, for $U=5\,eV$ (left panels)
  and $U=6\,eV$ (right panels) at fixed $J=0.7\,eV$.
  The vertical continue lines indicate the signature of the transition
  in the pressure-volume diagram, and the dotted vertical lines 
  indicate the boundary of the $\gamma$-$\alpha$ transition
  (which occurs only for $U\leq U_c\simeq 5.5\,eV$)
  according to the equal-area construction~\cite{Landau}.
}
\label{figure4}
\end{center}
\end{figure}

It is useful to examine how the local crossover induced
by the SOC
reflects on the quasi-particle renormalization weights and the on-site
configuration probabilities.
In the first panel of Fig.~\ref{figure4} 
are illustrated the averaged quasi-particle renormalization weights $Z$
of the $7/2$ and $5/2$ $f$-electrons --- which are 
significantly different
because of the spin-orbit effect.
As expected~\cite{Taylor05}, the $f$-electrons are correlated
($Z$ is significantly smaller than $1$)
even in the $\alpha$ phase, 
and the two $Z$'s monotonically decrease by increasing
the volume at higher pressures. Nevertheless, they 
develop a qualitatively different behaviour at the crossover point.
While the $5/2$ electrons undergo a clear crossover toward a
localized phase ``disentangled'' by the conduction electrons,
the $7/2$ electrons remain screened, but they rapidly disappear afterwards,
so that they are essentially absent in the $\gamma$ phase, see the second panel
of Fig.~\ref{figure4}.
As shown by the $f$ configuration probabilities
in the third panel of Fig.~\ref{figure4}, 
the SOC
speeds up the formation of the 4$f^1$ local moment.
In other words, the SOC acts as a ``catalyst'', which
favors the disentanglement between the 4$f$ electrons and the conduction
electrons.

In conclusion, we have performed first principle calculations
on Cerium using a new efficient implementation of the LDA+GA method.
For the physical value of $U$ in Cerium, $U\simeq 6\,eV$~\cite{McMahan98,Keller01},
a sharp crossover is observed in many physical quantities around the volume where
the bulk-modulus 
is minimum.
This finding is robust against changes in $U$, but the details are different.
For $U<U_c$, at $T=0$, there is a first-order transition
at a given negative value of the pressure,
while for $U>U_c$ this transition becomes a sharp crossover.
At $U_c$ there is a second-order quantum critical point in the phase diagram.
Our estimate for the critical interaction strength,
$U_c\simeq 5.5\,eV$, is very close to the physical value of the
interaction strength in Cerium.
This finding suggests that elemental Cerium is a critical element,
consistently with the experiments~\cite{Lashley}.
Our results demonstrate the importance of the SOC
for the volume collapse in Cerium, which is neatly captured by the rapid
variation of the entanglement
entropy of the $f$-electrons in the region
around the minimum of the bulk-modulus. 
In the $\alpha$ phase, at small $V$, the $f$-levels are
strongly hybridized with the conduction electrons, and the quasiparticle
weights and the pressure are only weakly-dependent on the
spin-orbit interaction. In this regime the system effectively behaves
as if the $f$-level degeneracy was of the order of magnitude of $14$.
In the $\gamma$ phase, at large $V$,
the spin-orbit splitting becomes more important, and substantially 
reduces the effective $f$-level degeneracy. 
The fact that the quasi-particle weight 
in the $\gamma$ phase is much smaller when the SOC is taken into account,
see Fig.~\ref{figure4}, can be interpreted as a consequence of the
above-mentioned reduction of effective $f$-level degeneracy
--- a well known effect in the theory of the single-ion Kondo impurity.
As in the early Kondo volume collapse theory~\cite{Allen82},
$\partial Z/\partial V$ contributes to the pressure. However,
some qualitative
features of our solution, such as 
the form of the pressure-volume phase diagram, show that other
physical elements,
such as the changes in the charge density induced by the correlations,
have to also be included in realistic theories of this material.

\section{CONTRIBUTIONS}

N.L. and Y.X.Y. co-developed the GA code
and analyzed the data.
N.L. developed the theoretical explanation of the 
iso-structural transition,
wrote the manuscript and carried out
the LDA+DMFT benchmark calculations.
Y.X.Y., C.Z.W. and K.M.H. initiated the project and carried out the LDA+GA calculations.
Y.X.Y. edited the figures,
and coded the interface between Wien2k and the GA solver,
which was constructed on the basis of the LDA+DMFT interface
developed by K.H..
K.H. developed the DMFT code.
G.K. and J.S. proposed the project
and supervised the research.
All the authors provided fundamental insights, and contributed
to improve the manuscript.
\\
\\
{\bf Corresponding author:} Y.X.Yao; ykent@iastate.edu

\section{ACKNOWLEDGMENTS}

N.L. and Y.X.Y. thank XiaoYu Deng 
and Robert McQueeney for useful discussion.
The collaboration was  supported by the U.S. Department of Energy
through the Computational Materials and Chemical Sciences Network CMSCN.
Research at Ames Laboratory supported by the U.S.
Department of Energy, Office of Basic Energy Sciences,
Division of Materials Sciences and Engineering. 
Ames Laboratory is operated for the U.S. Department of Energy
by Iowa State University under Contract No. DE-AC02-07CH11358.

\newpage
~
\newpage

\section{SUPPLEMENTARY MATERIAL}

\section{Methods}

In our LDA+GA code the LDA calculations are performed by Wien2k~\cite{w2k_S},
which is an all-electron DFT package based on the full-potential (linearized) 
augmented plane-wave ((L)APW)+local orbitals (lo) method.
A key advantage of this package is that it is one among the most accurate
schemes for band-structure calculations,  
and it is free of any numerical error due to the frozen-core pseudopotentials or from
the downfolding, which have been used in previous implementations of the LDA+GA method.

Our new LDA+GA implementation, to be described in a longer publication~[Y.Y.X. et al.],
is patterned after the LDA+DMFT work of Ref.~\cite{Haule10}. 
Using the same interface, basis set, projectors onto the correlated orbitals,
enables a meaningful comparison between the two methods.
We have employed a new numerical implementation of the
GA solver~[L.N. et. al, Strand Hugo U.R. et. al],
which further improves the method previously proposed in Ref.~\cite{Gmethod_S},
and is a generalization of earlier formulations of the
GA~\cite{Michele,Kondo,mybil,Deng_LDA+Gutz_S,Gebhard98}.

\section{LDA+DMFT benchmark calculations}

The purpose of this section is to benchmark our LDA+GA calculations 
within the LDA+DMFT method, using the Continuous Time Quantum Monte Carlo 
method (CTQMC)~\cite{ctqmc} --- which is 
numerically-exact --- as the impurity solver. 
We use the implementation or Ref~\cite{haule_ctmqc}.

Our DMFT calculations are all performed 
at $T=58\,K$, while the GA calculations are
done at zero temperature.

\begin{figure}
\begin{center}
\includegraphics[width=8.2cm]{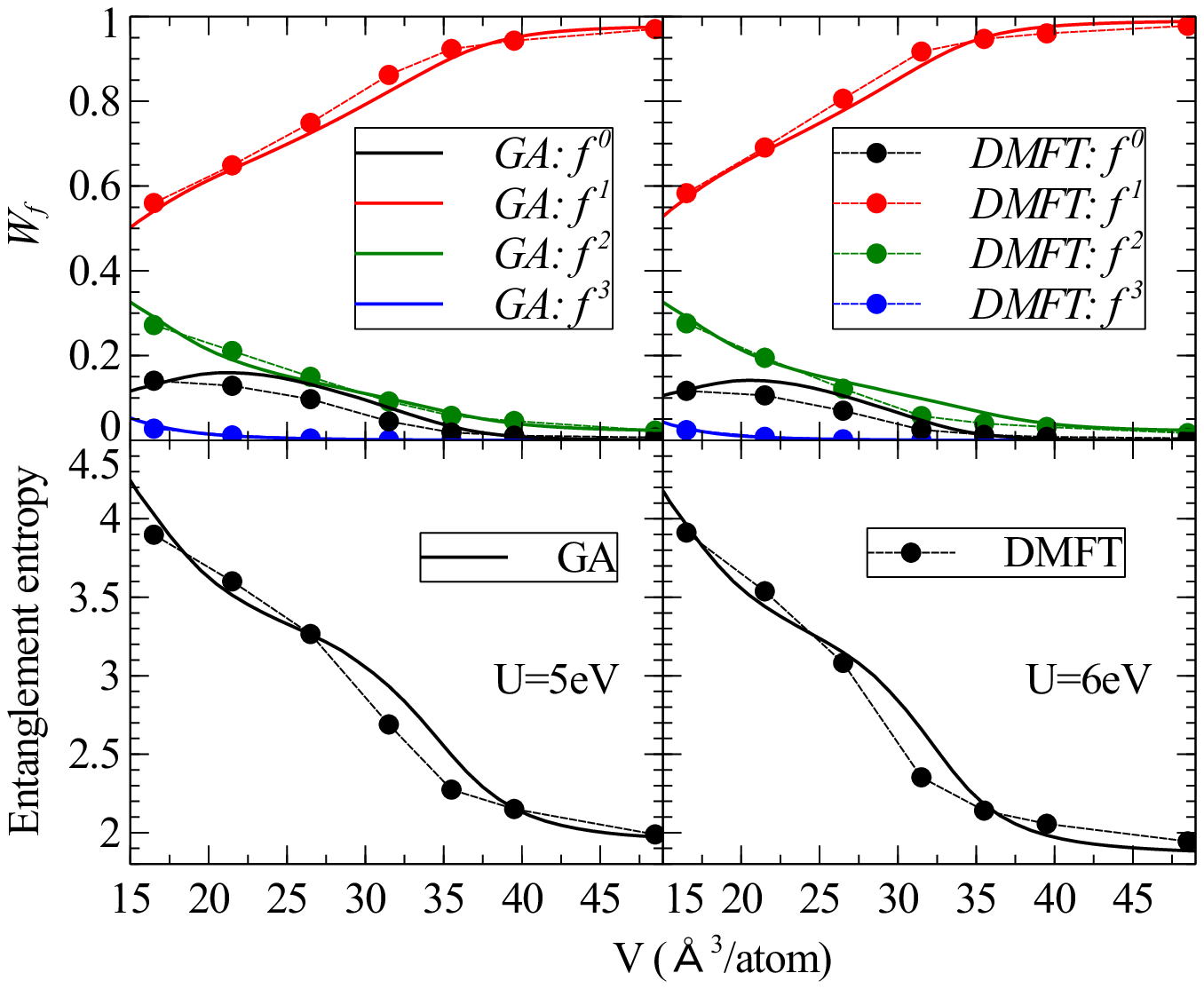}
\caption{Local configuration weights (upper panels) and
  $f$ entanglement entropy (lower panels).
  Comparison between LDA+GA (lines) and LDA+DMFT (dots) results
  for $U=5\,eV$ (left panels)
  and $U=6\,eV$ (right panels) at fixed $J=0.7\,eV$.
}
\label{figure5}
\end{center}
\end{figure}

\begin{figure}
\begin{center}
\includegraphics[width=8.2cm]{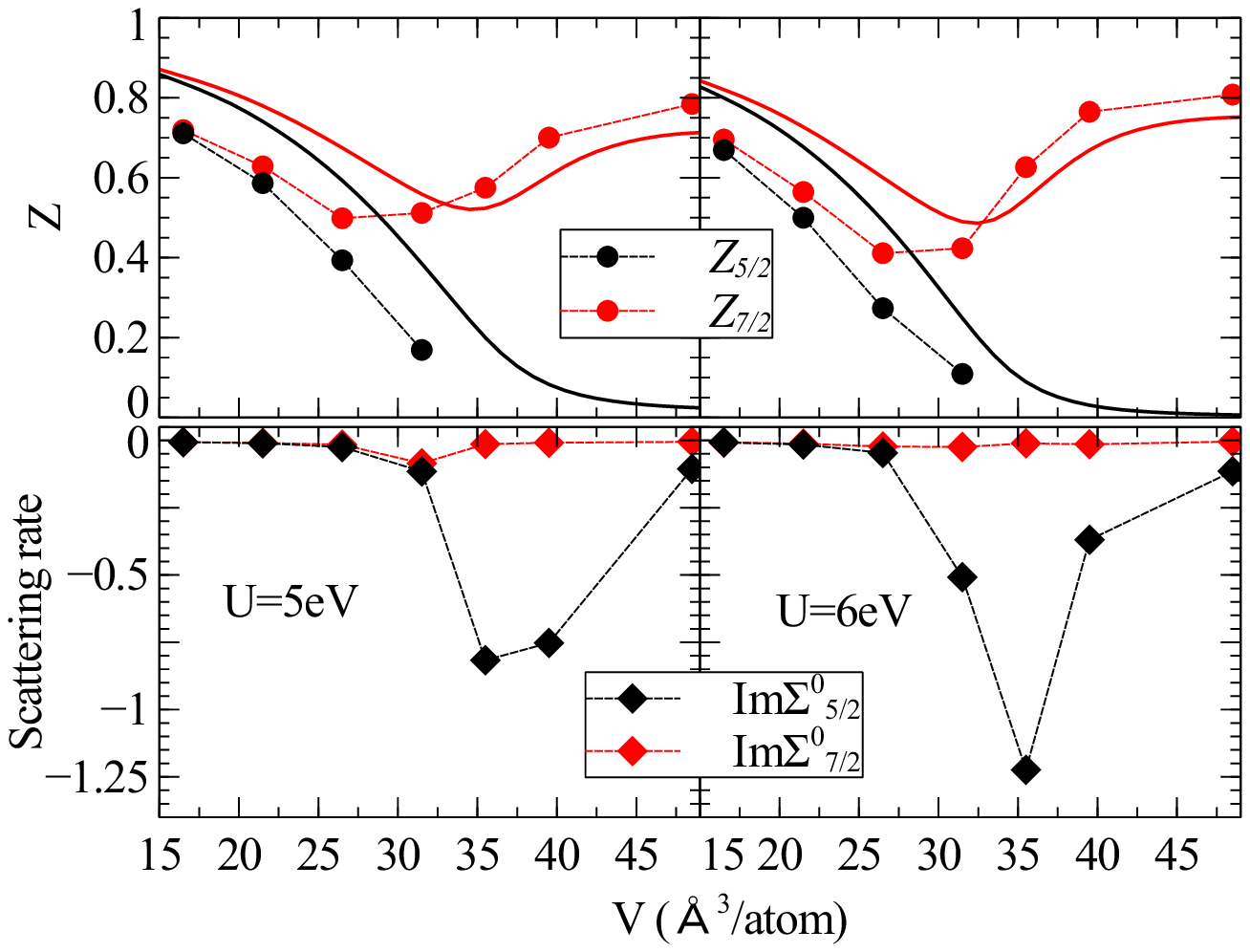}
\caption{Quasi-particle renormalization weights
  of the $7/2$ and $5/2$ $f$-electrons  (upper panels),
  and imaginary part of the self-energy at the Fermi level (lower panels).
  Comparison between LDA+GA (lines) and LDA+DMFT results (dots)
  for $U=5\,eV$ (left panels)
  and $U=6\,eV$ (right panels) at fixed $J=0.7\,eV$.
}
\label{figure6}
\end{center}
\end{figure}

In Fig.~\ref{figure5} are shown the local configuration probabilities (upper panels)
and the $f$ entanglement entropy (lower panels) as a function of the volume, 
for two different values of $U$.
The agreement between the two methods is indeed very good.
In particular, 
we point out that the $f$ local crossover, which is indicated by the rapid change
of the corresponding entanglement entropy,
is clearly visible also within the LDA+DMFT method.

In Fig.~\ref{figure6} it is shown the evolution of 
the quasi-particle renormalization weights as a function of the volume (upper panels)
and the imaginary part of the self-energy at the Fermi level (lower panels). 
Note that in the $\gamma$ phase, i.e., after the local crossover,
the $5/2$ electrons are no longer coherent. 
For this reason, at the temperature at which out DMFT calculations have been
carried out, $Z_{5/2}$ is not well defined at large volumes.


\begin{thebibliography}{99}
\expandafter\ifx\csname url\endcsname\relax
  \def\url#1{\texttt{#1}}\fi
\expandafter\ifx\csname urlprefix\endcsname\relax\def\urlprefix{URL }\fi
\providecommand{\bibinfo}[2]{#2}
\providecommand{\eprint}[2][]{\url{#2}}

\bibitem{Gschneidner78}
\bibinfo{author}{Koskenmaki, D.~C.} \& \bibinfo{author}{Jr., K. A.~G.}
\newblock \bibinfo{title}{Chapter 4 cerium}.
\newblock In \bibinfo{editor}{Karl A.~Gschneidner, J.} \&
  \bibinfo{editor}{Eyring, L.} (eds.) \emph{\bibinfo{booktitle}{Metals}},
  vol.~\bibinfo{volume}{1} of \emph{\bibinfo{series}{Handbook on the Physics
  and Chemistry of Rare Earths}}, \bibinfo{pages}{337 -- 377}
  (\bibinfo{publisher}{Elsevier}, \bibinfo{year}{1978}).

\bibitem{Lawson49}
\bibinfo{author}{Lawson, A.~W.} \& \bibinfo{author}{Tang, T.-Y.}
\newblock \bibinfo{title}{Concerning the high pressure allotropic modification
  of cerium}.
\newblock \emph{\bibinfo{journal}{Phys. Rev.}} \textbf{\bibinfo{volume}{76}},
  \bibinfo{pages}{301--302} (\bibinfo{year}{1949}).

\bibitem{Allen82}
\bibinfo{author}{Allen, J.~W.} \& \bibinfo{author}{Martin, R.~M.}
\newblock \bibinfo{title}{Kondo volume collapse and the $\gamma \rightarrow
  \alpha$ transition in cerium}.
\newblock \emph{\bibinfo{journal}{Phys. Rev. Lett.}}
  \textbf{\bibinfo{volume}{49}}, \bibinfo{pages}{1106--1110}
  (\bibinfo{year}{1982}).

\bibitem{Lavagna82}
\bibinfo{author}{Lavagna, M.}, \bibinfo{author}{Lacroix, C.} \&
  \bibinfo{author}{Cyrot, M.}
\newblock \bibinfo{title}{Volume collapse in the kondo lattice}.
\newblock \emph{\bibinfo{journal}{Phys. Lett. A}}
  \textbf{\bibinfo{volume}{90}}, \bibinfo{pages}{210 -- 212}
  (\bibinfo{year}{1982}).

\bibitem{Johansson74}
\bibinfo{author}{Johansson, B.}
\newblock \bibinfo{title}{The $\alpha$-$\gamma$ transition in cerium is a mott
  transition}.
\newblock \emph{\bibinfo{journal}{Philos. Mag.}} \textbf{\bibinfo{volume}{30}},
  \bibinfo{pages}{469} (\bibinfo{year}{1974}).

\bibitem{Wieliczka}
\bibinfo{author}{Wieliczka, D.~M.}, \bibinfo{author}{Olson, C.~G.} \&
  \bibinfo{author}{Lynch, D.~W.}
\newblock \bibinfo{title}{Valence-band photoemission in la and pr: Connections
  with the Ce problem}.
\newblock \emph{\bibinfo{journal}{Phys. Rev. Lett.}}
  \textbf{\bibinfo{volume}{52}}, \bibinfo{pages}{2180--2182}
  (\bibinfo{year}{1984}).

\bibitem{Weschke}
\bibinfo{author}{Weschke, E.} \emph{et~al.}
\newblock \bibinfo{title}{Surface and bulk electronic structure of Ce metal
  studied by high-resolution resonant photoemission}.
\newblock \emph{\bibinfo{journal}{Phys. Rev. B}} \textbf{\bibinfo{volume}{44}},
  \bibinfo{pages}{8304--8307} (\bibinfo{year}{1991}).

\bibitem{Patthey}
\bibinfo{author}{Patthey, F.}, \bibinfo{author}{Delley, B.},
  \bibinfo{author}{Schneider, W.~D.} \& \bibinfo{author}{Baer, Y.}
\newblock \bibinfo{title}{Low-energy excitations in $\alpha$- and $\gamma$-ce
  observed by photoemission}.
\newblock \emph{\bibinfo{journal}{Phys. Rev. Lett.}}
  \textbf{\bibinfo{volume}{55}}, \bibinfo{pages}{1518--1521}
  (\bibinfo{year}{1985}).

\bibitem{Keller01}
\bibinfo{author}{Z\"olfl, M.~B.}, \bibinfo{author}{Nekrasov, I.~A.},
  \bibinfo{author}{Pruschke, T.}, \bibinfo{author}{Anisimov, V.~I.} \&
  \bibinfo{author}{Keller, J.}
\newblock \bibinfo{title}{Spectral and magnetic properties of $\alpha$- and
  $\gamma$-Ce from dynamical mean-field theory and local density
  approximation}.
\newblock \emph{\bibinfo{journal}{Phys. Rev. Lett.}}
  \textbf{\bibinfo{volume}{87}}, \bibinfo{pages}{276403}
  (\bibinfo{year}{2001}).

\bibitem{Held01}
\bibinfo{author}{Held, K.}, \bibinfo{author}{McMahan, A.~K.} \&
  \bibinfo{author}{Scalettar, R.~T.}
\newblock \bibinfo{title}{Cerium volume collapse: Results from the merger of
  dynamical mean-field theory and local density approximation}.
\newblock \emph{\bibinfo{journal}{Phys. Rev. Lett.}}
  \textbf{\bibinfo{volume}{87}}, \bibinfo{pages}{276404}
  (\bibinfo{year}{2001}).

\bibitem{McMahan}
\bibinfo{author}{McMahan, A.~K.}, \bibinfo{author}{Held, K.} \&
  \bibinfo{author}{Scalettar, R.~T.}
\newblock \bibinfo{title}{Thermodynamic and spectral properties of compressed
  Ce calculated using a combined local-density approximation and dynamical
  mean-field theory}.
\newblock \emph{\bibinfo{journal}{Phys. Rev. B}} \textbf{\bibinfo{volume}{67}},
  \bibinfo{pages}{075108} (\bibinfo{year}{2003}).

\bibitem{LDA+U+DMFT}
\bibinfo{author}{Kotliar, G.} \emph{et~al.}
\newblock \bibinfo{title}{Electronic structure calculations with dynamical
  mean-field theory}.
\newblock \emph{\bibinfo{journal}{Rev. Mod. Phys.}}
  \textbf{\bibinfo{volume}{78}}, \bibinfo{pages}{865} (\bibinfo{year}{2006}).

\bibitem{Gabi05}
\bibinfo{author}{Haule, K.}, \bibinfo{author}{Oudovenko, V.},
  \bibinfo{author}{Savrasov, S.~Y.} \& \bibinfo{author}{Kotliar, G.}
\newblock \bibinfo{title}{The $\alpha\rightarrow\gamma$ transition in ce: A
  theoretical view from optical spectroscopy}.
\newblock \emph{\bibinfo{journal}{Phys. Rev. Lett.}}
  \textbf{\bibinfo{volume}{94}}, \bibinfo{pages}{036401}
  (\bibinfo{year}{2005}).

\bibitem{Gabi06}
\bibinfo{author}{Kotliar, G.} \emph{et~al.}
\newblock \bibinfo{title}{Electronic structure calculations with dynamical
  mean-field theory}.
\newblock \emph{\bibinfo{journal}{Rev. Mod. Phys.}}
  \textbf{\bibinfo{volume}{78}}, \bibinfo{pages}{865--951}
  (\bibinfo{year}{2006}).

\bibitem{Moore09}
\bibinfo{author}{Moore, K.~T.} \& \bibinfo{author}{van~der Laan, G.}
\newblock \bibinfo{title}{Nature of the $5f$ states in actinide metals}.
\newblock \emph{\bibinfo{journal}{Rev. Mod. Phys.}}
  \textbf{\bibinfo{volume}{81}}, \bibinfo{pages}{235--298}
  (\bibinfo{year}{2009}).

\bibitem{deMedici}
\bibinfo{author}{de' Medici, L.}, \bibinfo{author}{Georges, A.},
  \bibinfo{author}{Kotliar, G.} \& \bibinfo{author}{Biermann, S.}
\newblock \bibinfo{title}{Mott transition and kondo screening in $f$-electron
  metals}.
\newblock \emph{\bibinfo{journal}{Phys. Rev. Lett.}}
  \textbf{\bibinfo{volume}{95}}, \bibinfo{pages}{066402}
  (\bibinfo{year}{2005}).

\bibitem{Georges06}
\bibinfo{author}{Amadon, B.}, \bibinfo{author}{Biermann, S.},
  \bibinfo{author}{Georges, A.} \& \bibinfo{author}{Aryasetiawan, F.}
\newblock \bibinfo{title}{The $\alpha$-$\gamma$ transition of cerium is entropy
  driven}.
\newblock \emph{\bibinfo{journal}{Phys. Rev. Lett.}}
  \textbf{\bibinfo{volume}{96}}, \bibinfo{pages}{066402}
  (\bibinfo{year}{2006}).

\bibitem{Scheffler12}
\bibinfo{author}{Casadei, M.}, \bibinfo{author}{Ren, X.},
  \bibinfo{author}{Rinke, P.}, \bibinfo{author}{Rubio, A.} \&
  \bibinfo{author}{Scheffler, M.}
\newblock \bibinfo{title}{Density-functional theory for $f$-electron systems:
  The $\alpha$-$\gamma$ phase transition in cerium}.
\newblock \emph{\bibinfo{journal}{Phys. Rev. Lett.}}
  \textbf{\bibinfo{volume}{109}}, \bibinfo{pages}{146402}
  (\bibinfo{year}{2012}).

\bibitem{Gutz65}
\bibinfo{author}{Gutzwiller, M.~C.}
\newblock \bibinfo{title}{Correlation of electrons in a narrow $s$ band}.
\newblock \emph{\bibinfo{journal}{Phys. Rev.}} \textbf{\bibinfo{volume}{137}},
  \bibinfo{pages}{A1726--A1735} (\bibinfo{year}{1965}).

\bibitem{Zein}
\bibinfo{author}{Zein, N.~E.}
\newblock \bibinfo{title}{Correlation energy functionals for \textit{ab initio}
  calculations: Application to transition metals}.
\newblock \emph{\bibinfo{journal}{Phys. Rev. B}} \textbf{\bibinfo{volume}{52}},
  \bibinfo{pages}{11813--11824} (\bibinfo{year}{1995}).

\bibitem{Deng_LDA+Gutz}
\bibinfo{author}{Deng, X.}, \bibinfo{author}{Wang, L.}, \bibinfo{author}{Dai,
  X.} \& \bibinfo{author}{Fang, Z.}
\newblock \bibinfo{title}{Local density approximation combined with gutzwiller
  method for correlated electron systems: Formalism and applications}.
\newblock \emph{\bibinfo{journal}{Phys. Rev. B}} \textbf{\bibinfo{volume}{79}},
  \bibinfo{pages}{075114} (\bibinfo{year}{2009}).

\bibitem{Ho}
\bibinfo{author}{Ho, K.~M.}, \bibinfo{author}{Schmalian, J.} \&
  \bibinfo{author}{Wang, C.~Z.}
\newblock \bibinfo{title}{Gutzwiller density functional theory for correlated
  electron systems}.
\newblock \emph{\bibinfo{journal}{Phys. Rev. B}} \textbf{\bibinfo{volume}{77}},
  \bibinfo{pages}{073101} (\bibinfo{year}{2008}).

\bibitem{Gmethod}
\bibinfo{author}{Lanat\`a, N.}, \bibinfo{author}{Strand, H. U.~R.},
  \bibinfo{author}{Dai, X.} \& \bibinfo{author}{Hellsing, B.}
\newblock \bibinfo{title}{Efficient implementation of the gutzwiller
  variational method}.
\newblock \emph{\bibinfo{journal}{Phys. Rev. B}} \textbf{\bibinfo{volume}{85}},
  \bibinfo{pages}{035133} (\bibinfo{year}{2012}).

\bibitem{w2k}
\bibinfo{author}{Blaha, P.}, \bibinfo{author}{Schwarz, K.},
  \bibinfo{author}{Madsen, G.}, \bibinfo{author}{Kvasnicka, D.} \&
  \bibinfo{author}{Luitz, J.}
\newblock \emph{\bibinfo{journal}{An augmented plane wave plus local orbitals
  program for calculating crystal properties. University of Technology,
  Vienna}}  (\bibinfo{year}{2001}).

\bibitem{cowan}
\bibinfo{author}{Cowan, R.~D.}
\newblock \emph{\bibinfo{title}{The Theory of Atomic Structure and Spectra}}
  (\bibinfo{publisher}{University of California Press, Berkeley},
  \bibinfo{year}{1981}).

\bibitem{CePV74}
\bibinfo{author}{Ellinger, F.~H.} \& \bibinfo{author}{Zachariasen, W.~H.}
\newblock \bibinfo{title}{Structure of cerium metal at high pressure}.
\newblock \emph{\bibinfo{journal}{Phys. Rev. Lett.}}
  \textbf{\bibinfo{volume}{32}}, \bibinfo{pages}{773--774}
  (\bibinfo{year}{1974}).

\bibitem{CePV77}
\bibinfo{author}{Zachariasen, W.~H.} \& \bibinfo{author}{Ellinger, F.~H.}
\newblock \bibinfo{title}{{The crystal structures of cerium metal at high
  pressure}}.
\newblock \emph{\bibinfo{journal}{Acta Crystallogr. Sec. A}}
  \textbf{\bibinfo{volume}{33}}, \bibinfo{pages}{155--160}
  (\bibinfo{year}{1977}).

\bibitem{CePV85}
\bibinfo{author}{Olsen, J.}, \bibinfo{author}{Gerward, L.},
  \bibinfo{author}{Benedict, U.} \& \bibinfo{author}{Iti\'{e}, J.-P.}
\newblock \bibinfo{title}{The crystal structure and the equation of state of
  cerium metal in the pressure range 0-–46 GPa}.
\newblock \emph{\bibinfo{journal}{Physica {B+C}}}
  \textbf{\bibinfo{volume}{133}}, \bibinfo{pages}{129--137}
  (\bibinfo{year}{1985}).

\bibitem{Lipp12}
\bibinfo{author}{Lipp, M.~J.} \emph{et~al.}
\newblock \bibinfo{title}{X-ray emission spectroscopy of cerium across the
  $\gamma$-$\alpha$ volume collapse transition}.
\newblock \emph{\bibinfo{journal}{Phys. Rev. Lett.}}
  \textbf{\bibinfo{volume}{109}}, \bibinfo{pages}{195705}
  (\bibinfo{year}{2012}).

\bibitem{McMahan98}
\bibinfo{author}{McMahan, A.}, \bibinfo{author}{Huscroft, C.},
  \bibinfo{author}{Scalettar, R.} \& \bibinfo{author}{Pollock, E.}
\newblock \bibinfo{title}{Volume-collapse transitions in the rare earth
  metals}.
\newblock \emph{\bibinfo{journal}{Journal of Computer-Aided Materials Design}}
  \textbf{\bibinfo{volume}{5}}, \bibinfo{pages}{131--162}
  (\bibinfo{year}{1998}).

\bibitem{Landau}
\bibinfo{author}{Landau, L.~D.} \& \bibinfo{author}{Lifshitz, E.~M.}
\newblock \emph{\bibinfo{title}{Statistical Physics}}
  (\bibinfo{publisher}{Peramon, London}, \bibinfo{year}{1958}).

\bibitem{Taylor05}
\bibinfo{author}{Murani, A.~P.}, \bibinfo{author}{Levett, S.~J.} \&
  \bibinfo{author}{Taylor, J.~W.}
\newblock \bibinfo{title}{Magnetic form factor of $\alpha$-ce: Towards
  understanding the magnetism of cerium}.
\newblock \emph{\bibinfo{journal}{Phys. Rev. Lett.}}
  \textbf{\bibinfo{volume}{95}}, \bibinfo{pages}{256403}
  (\bibinfo{year}{2005}).

\bibitem{Lashley}
\bibinfo{author}{Lashley, J.~C.} \emph{et~al.}
\newblock \bibinfo{title}{Tricritical phenomena at the $\gamma \rightarrow
  \alpha$ transition in Ce$_{0.9-x}$La$_{x}$Th$_{0.1}$ alloys}.
\newblock \emph{\bibinfo{journal}{Phys. Rev. Lett.}}
  \textbf{\bibinfo{volume}{97}}, \bibinfo{pages}{235701}
  (\bibinfo{year}{2006}).

\end{thebibliography}

\begin{thebibliography}{99}
\expandafter\ifx\csname url\endcsname\relax
  \def\url#1{\texttt{#1}}\fi
\expandafter\ifx\csname urlprefix\endcsname\relax\def\urlprefix{URL }\fi
\providecommand{\bibinfo}[2]{#2}
\providecommand{\eprint}[2][]{\url{#2}}

\bibitem{w2k_S}
\bibinfo{author}{Blaha, P.}, \bibinfo{author}{Schwarz, K.},
  \bibinfo{author}{Madsen, G.}, \bibinfo{author}{Kvasnicka, D.} \&
  \bibinfo{author}{Luitz, J.}
\newblock \emph{\bibinfo{journal}{An augmented plane wave plus local orbitals
  program for calculating crystal properties. University of Technology,
  Vienna}}  (\bibinfo{year}{2001}).

\bibitem{Haule10}
\bibinfo{author}{Haule, K.}, \bibinfo{author}{Yee, C.-H.} \&
  \bibinfo{author}{Kim, K.}
\newblock \bibinfo{title}{Dynamical mean-field theory within the full-potential
  methods: Electronic structure of CeIrIn$_{5}$, CeCoIn$_{5}$, and
  CeRhIn$_{5}$}.
\newblock \emph{\bibinfo{journal}{Phys. Rev. B}} \textbf{\bibinfo{volume}{81}},
  \bibinfo{pages}{195107} (\bibinfo{year}{2010}).

\bibitem{Gmethod_S}
\bibinfo{author}{Lanat\`a, N.}, \bibinfo{author}{Strand, H. U.~R.},
  \bibinfo{author}{Dai, X.} \& \bibinfo{author}{Hellsing, B.}
\newblock \bibinfo{title}{Efficient implementation of the gutzwiller
  variational method}.
\newblock \emph{\bibinfo{journal}{Phys. Rev. B}} \textbf{\bibinfo{volume}{85}},
  \bibinfo{pages}{035133} (\bibinfo{year}{2012}).

\bibitem{Michele}
\bibinfo{author}{Fabrizio, M.}
\newblock \bibinfo{title}{Gutzwiller description of non-magnetic {Mott}
  insulators: {Dimer} lattice model}.
\newblock \emph{\bibinfo{journal}{Phys. Rev. B}} \textbf{\bibinfo{volume}{76}},
  \bibinfo{pages}{165110} (\bibinfo{year}{2007}).

\bibitem{Kondo}
\bibinfo{author}{Lanat\`{a}, N.}, \bibinfo{author}{Barone, P.} \&
  \bibinfo{author}{Fabrizio, M.}
\newblock \bibinfo{title}{Fermi-surface evolution across the magnetic phase
  transition in the {Kondo} lattice model}.
\newblock \emph{\bibinfo{journal}{Phys. Rev. B}} \textbf{\bibinfo{volume}{78}},
  \bibinfo{pages}{155127} (\bibinfo{year}{2008}).

\bibitem{mybil}
\bibinfo{author}{Lanat\`a, N.}, \bibinfo{author}{Barone, P.} \&
  \bibinfo{author}{Fabrizio, M.}
\newblock \bibinfo{title}{Superconductivity in the doped bilayer hubbard
  model}.
\newblock \emph{\bibinfo{journal}{Phys. Rev. B}} \textbf{\bibinfo{volume}{80}},
  \bibinfo{pages}{224524} (\bibinfo{year}{2009}).

\bibitem{Deng_LDA+Gutz_S}
\bibinfo{author}{Deng, X.}, \bibinfo{author}{Wang, L.}, \bibinfo{author}{Dai,
  X.} \& \bibinfo{author}{Fang, Z.}
\newblock \bibinfo{title}{Local density approximation combined with gutzwiller
  method for correlated electron systems: Formalism and applications}.
\newblock \emph{\bibinfo{journal}{Phys. Rev. B}} \textbf{\bibinfo{volume}{79}},
  \bibinfo{pages}{075114} (\bibinfo{year}{2009}).

\bibitem{Gebhard98}
\bibinfo{author}{B\"unemann, J.}, \bibinfo{author}{Weber, W.} \&
  \bibinfo{author}{Gebhard, F.}
\newblock \bibinfo{title}{Multiband gutzwiller wave functions for general
  on-site interactions}.
\newblock \emph{\bibinfo{journal}{Phys. Rev. B}} \textbf{\bibinfo{volume}{57}},
  \bibinfo{pages}{6896--6916} (\bibinfo{year}{1998}).

\bibitem{ctqmc}
\bibinfo{author}{Werner, P.}, \bibinfo{author}{Comanac, A.},
  \bibinfo{author}{de' Medici, L.}, \bibinfo{author}{Troyer, M.} \&
  \bibinfo{author}{Millis, A.~J.}
\newblock \bibinfo{title}{Continuous-time solver for quantum impurity models}.
\newblock \emph{\bibinfo{journal}{Phys. Rev. Lett.}}
  \textbf{\bibinfo{volume}{97}}, \bibinfo{pages}{076405}
  (\bibinfo{year}{2006}).

\bibitem{haule_ctmqc}
\bibinfo{author}{Haule, K.}
\newblock \bibinfo{title}{Quantum monte carlo impurity solver for cluster
  dynamical mean-field theory and electronic structure calculations with
  adjustable cluster base}.
\newblock \emph{\bibinfo{journal}{Phys. Rev. B}} \textbf{\bibinfo{volume}{75}},
  \bibinfo{pages}{155113} (\bibinfo{year}{2007}).

\end{thebibliography}

\end{document}